\documentclass{raa}
\usepackage{natbib,times}
\usepackage{lscape}

\begin{document}

\title{Old stellar population synthesis: New age and mass estimates for
Mayall II = G1}

\volnopage{Vol.0 (200x) No.0, 000--000}      
\setcounter{page}{1}           

\author{Jun Ma
\inst{1}\mailto{}
\and Richard de Grijs\inst{2,1}
\and Zhou Fan\inst{1,3}
\and Soo-Chang Rey\inst{4}
\and Zhenyu Wu\inst{1}
\and Xu Zhou\inst{1}
\and Jianghua Wu\inst{1}
\and Zhaoji Jiang\inst{1}
\and Jiansheng Chen\inst{1}
\and Kyungsook Lee\inst{4}
\and S. T. Sohn\inst{5,6}
}

\institute{National Astronomical Observatories, Chinese
Academy of Sciences, Beijing, 100012, P. R. China\\
\email{majun@vega.bac.pku.edu.cn}
\and
Department of Physics \& Astronomy, The University of
Sheffield, Hicks Building, Hounsfield Road, Sheffield S3 7RH, UK\\
\and
Graduate University, Chinese Academy of
Sciences, Beijing, 100039, P. R. China\\
\and
Department of Astronomy and Space Science, Chungnam
National University, Daejeon 305-764, Korea\\
\and
Center for Space Astrophysics, Yonsei University, Seoul 120-749, Korea\\
\and
California Institute of Technology, MC 405-47, 1200
E.  California Boulevard, Pasadena, CA 91125\\
}

\date{Received~~2001 month day; accepted~~2001~~month day}

\abstract{
Mayall II = G1 is one of the most luminous globular clusters (GCs) in
M31. Here, we determine its age and mass by comparing multicolor
photometry with theoretical stellar population synthesis models. Based
on far- and near-ultraviolet {\sl GALEX} photometry, broad-band
$UBVRI$, and infrared $JHK_{\rm s}$ 2MASS data, we construct the most
extensive spectral energy distribution of G1 to date, spanning the
wavelength range from 1538 to 20,000{\AA}. A quantitative comparison
with a variety of simple stellar population (SSP) models yields a mean
age that is consistent with G1 being among the oldest building blocks
of M31 and having formed within $\sim$1.7 Gyr after the Big Bang.
Irrespective of the SSP model or stellar initial mass function
adopted, the resulting mass estimates (of order $10^7 M_\odot$)
indicate that G1 is one of the most massive GCs in the Local Group.
However, we speculate that the cluster's exceptionally high mass
suggests that it may not be a genuine GC. We also derive that G1 may
contain, on average, $(1.65\pm0.63)\times10^2 L_\odot$
far-ultraviolet-bright, hot, extreme horizontal-branch stars,
depending on the SSP model adopted. On a generic level, we demonstrate
that extensive multi-passband photometry coupled with SSP analysis
enables one to obtain age estimates for old SSPs to a similar accuracy
as from integrated spectroscopy or resolved stellar photometry,
provided that some of the free parameters can be constrained
independently.
}

\authorrunning{Ma et al.}
\titlerunning{New age and mass estimates of G1}

\maketitle

\section{Introduction}

Globular clusters (GCs) are among the oldest bound stellar systems
in the Universe, and they thus provide a fossil record of the
earliest stages of galaxy formation and evolution. GCs are
internally homogeneous in age and metallicity, i.e. they are simple
stellar systems composed of coeval stellar populations. In addition,
GCs are the oldest systems in our own and other galaxies for which
we can estimate reasonably reliable ages (and realistic
uncertainties); they can thus independently provide vitally
important information regarding the minimum age of the Universe. In
a detailed study of 17 Galactic GCs, \cite{chaboyer98} used the
improved {\sl Hipparcos} parallaxes having just become available at
that time to determine updated distances, and hence improved ages,
of their GC sample. They concluded that the mean age of the oldest
GCs is $11.5 \pm 1.3$ Gyr, although their age histogram (their fig.
2) shows a tail toward ages as old as $\sim 16$ Gyr.
\cite{gratton03} obtain improved ages (and distances) for three
Galactic GCs, NGC 6397, NGC 6752, and 47 Tuc, and conclude that the
age of the oldest GCs is $13.4 \pm 0.8$ (random) $\pm 0.6$
(systematic) Gyr, in good agreement with the 3-year results from the
{\sl Wilkinson Microwave Anisotropy Probe (WMAP)}. This led them to
suggest that the oldest Galactic GCs formed within the first 1.7 Gyr
after the Big Bang, at the $1\sigma$ confidence level. We note that
this is still fully compatible with the 5-year WMAP results
constraining the age of the Universe to $13.73 \pm 0.12$ Gyr
(Hinshaw et al. 2008). While the ages of the oldest GCs in the Galaxy are
now reasonably well determined, this is certainly not the case even
for our nearest large neighbour, the Andromeda galaxy (M31).

The most direct method for determining the age of a star cluster is by
means of individual stellar photometry, since the main-sequence
turn-off location is mostly affected by age (see, e.g., Puzia et al. 2002b, and
references therein). However, this method has only been
applied to Galactic GCs and to GCs associated with the Milky Way's
satellites (e.g., Riich 2001), in which individual stars can be
resolved and their photometry determined to satisfactory accuracy, to
a few magnitudes fainter than the main-sequence turn-offs. This is
difficult, if not impossible, to achieve even for GCs as close as
those associated with M31 (but see Brown et al. 2004), at a distance
of $D = 772 \pm 44$ kpc (e.g., Ribas et al. 2005).  Fortunately,
starting from the pioneering work of Tinsley (1968, 1972) and
\cite{SSB73}, evolutionary population synthesis modeling has become a
powerful tool to address many outstanding problems in astrophysics,
from determining the ages of star clusters to interpreting integrated
spectrophotometric observations of galaxies. Therefore, extragalactic
GC ages can, in general, also be inferred from composite colors and/or
integrated spectroscopy.

The evolution of GCs is usually modeled by means of the simple
stellar population (SSP) approximation. An SSP is a single
generation of coeval stars formed from the same progenitor molecular
cloud (thus implying a single metallicity), and governed by a given
stellar initial mass function (IMF). GCs are ideal templates to test
the compatibility between the population synthesis models and
reality. \cite{b00} compared the predicted SSP colors of three
stellar population synthesis models to the intrinsic broad-band
$UBVRIJHK$ colors of Galactic and M31 GCs, and concluded that the
best-fitting models match the cluster colors very well.
Subsequently, many authors have used SSP modeling to determine the
parameters of cluster populations. For instance, \cite{degrijs03a}
determined the ages and masses of star clusters in the fossil
starburst region B of M82 by comparing the observed cluster spectral
energy distributions (SEDs) with both the {\sc Starburst99} SSP
models (Leitherer et al. 1999) and those developed by \cite{bc00},
based on {\sl Hubble Space Telescope (HST)} observations from the
blue optical to the near-infrared (NIR) (see also de Grijs et al.
2003b, 2003c); \cite{bik03} and \cite{bastian05} estimated ages,
initial masses and extinction values for M51 star cluster candidates
by comparing the {\sc Starburst99} and the Frascati models
(Romaniello 1998) for instantaneous star formation with the observed
SEDs based on {\sl HST}/WFPC2 observations in six broad-band and two
narrow-band filters. \cite{ma06} and \cite{fan06} obtained age
estimates for M31 GCs by fitting theoretical stellar population
synthesis models (Bruzual \& Charlot 2003, henceforth BC03) to their
photometric measurements in a large number of intermediate- and
broad-band passbands from the optical to the NIR. Based on the same
method and models, \cite{ma07a} constrained the age of the M31 GC
S312, using multicolor photometry from the near-ultraviolet (NUV) to
the NIR, to $9.5^{+1.15}_{-0.99}$~Gyr. S312 is among the first
extragalactic GCs for which the age was estimated accurately using
main-sequence photometry, i.e., \cite{brown04} estimated the age of
S312 at $10^{+2.5}_{-1}$ Gyr by means of a quantitative comparison
with the isochrones of \cite{vandenberg06}. This was based on their
analysis of the cluster's color-magnitude diagram (CMD) below the
main-sequence turn-off using extremely deep images obtained with the
{\sl HST}/Advanced Camera for Surveys (ACS).

It is a common misconception that spectroscopic age estimates are
always much better than those based on broad-band photometry.
\cite{ssb04} recently showed convincingly that spectroscopic age
determinations are not necessarily better or more accurate than
photometrically obtained ages, at least in the age range from $\sim
100-500$ Myr. \cite{anders04} published a detailed theoretical
investigation of the accuracy of retrieved star cluster properties,
including their ages, based on sophisticated fits of SSP models to
observed broad-band SEDs spanning varying wavelength ranges. They
concluded that if one has access to as large a wavelength range as
possible, ideally including both ultraviolet and NIR data points,
the resulting age estimates are reasonably accurate, even for ages
as old as $\sim 10$ Gyr -- particularly if one or more of the other
free parameters (e.g., metallicity or extinction) can be constrained
reliably and independently (see Anders et al. 2004, e.g., their fig.
14). We will use this promising approach as our basic premise in
this paper.

Of the Local Group members, M31 is particularly important as it
provides a direct comparison with our own Galaxy. In addition, it
contains a large number of GCs and GC candidates, including 496
genuine, 367 probable, and 301 possible GCs (Kim et al. 2007). The M31 GC
system is among the extragalactic GC systems studied most often
(Harris 1991; Brodie \& Strader 2006). As one of the brightest M31 GCs, Mayall II
= G1 has attracted much scientific interest (see, e.g., Barmby et al. 2007;
Ma et al. 2007b, and references therein).

In this paper, we first determine the age and mass of G1 by comparing
observational SEDs (\S 2) with population synthesis models (\S 3). We
will use the lessons learned from studies of broad-band photometric SED
fits to minimize the associated uncertainties. We discuss our results
along the way, as appropriate, and provide a summary in \S 4.

\section{Ultraviolet, optical, and infrared observations of G1}

\subsection{Historical observations}

G1 was first detected by \cite{mayegg53} (their No. 2, and hence
referred to as Mayall II), who searched for nebulous objects
associated with M31 using a $6^{\circ}\times6^{\circ}$ Palomar 48-inch
Schmidt plate taken in 1948 and centered on M31. Subsequently,
\cite{sarg77} rediscovered the cluster (their No. 1, i.e., G1) based
on their survey of 29 plates associated with the general field of M31,
which had been obtained at the $f/2.7$ prime focus of the Kitt Peak
National Observatory's (KPNO) 4-m telescope. The cluster is located in
the halo of M31, at a projected distance of about 40 kpc from the
galaxy's nucleus (see Meylan et al. 2001).

\subsection{{\sl GALEX} ultraviolet, optical broad-band, and 2MASS NIR photometry}

Although the cluster is generally believed to be among the oldest GCs
in M31, to the best of our knowledge there is no CMD-based or
spectroscopic age estimate available in the literature to date. The
lack of a CMD-based age estimate is due to the challenges associated
with probing the cluster's CMD down to below the main-sequence
turn-off. The current deepest CMD of the cluster (Meylan et al. 2001)
does not reach these faint levels. Although both integrated and
spatially-resolved spectra of the cluster are available
(e.g., Huchra et al. 1991; Gebhardt et al. 2005; Cohen 2006), they have thus far not been used
to determine an age for G1. This may be partially due to the limited
wavelength range covered by most of these spectra, and the
difficulties one faces when trying to constrain ages in the regime
beyond $\sim 10$ Gyr (see below).

To constrain the age of G1 accurately, with the smallest uncertainty
allowed by the observational data, we use as many photometric data
points covering as large a wavelength range as possible.
\cite{kaviraj07} showed that the combination of far (FUV) and
near-ultraviolet photometry with optical observations in the standard
broad bands enables one to efficiently break the age-metallicity
degeneracy; \cite{wor94} showed that the age-metallicity degeneracy
associated with optical broad-band colors is $\Delta {\rm age}/\Delta
Z \sim 3/2$ (also see MacArthur et al. 2004). However, \cite{jong96}
showed that this degeneracy can be partially broken by adding NIR
photometry to the optical colors, which was recently supported by
\cite{wu05}. \cite{Cardiel03} found that inclusion of an infrared
passband can improve the predictive power of the stellar population
diagnostics by $\sim$30 times compared to using optical photometry
alone. Since NIR photometry is less sensitive to interstellar
extinction than the classical optical passbands, \cite{kbm02} and
\cite{puzia02a} also suggested that it provides useful complementary
information that can help to disentangle the age-metallicity
degeneracy (also see Galleti et al. 2004).

The M31 field was observed as part of the Nearby Galaxies Survey
(NGS) by the {\sl Galaxy Evolution Explorer} ({\sl GALEX}) in two
ultraviolet bands (see for details from Rey et al 2005, 2007).
\cite{rey06} published photometric data for 485 and 273 M31 GCs in
the {\sl GALEX} NUV and FUV bands, respectively. G1 was detected in
these two ultraviolet bands. The {\sl GALEX} photometric system is
calibrated to match the spectrophotometric AB system.

\cite{Sidney69} determined photo-electric photometry for 45 M31 GCs,
including G1, in the $UBV$ bands. Using CCD imaging from the KPNO 0.9m
telescope, \cite{rhh94} published integrated $BVR$ magnitudes and
color indices for 41 GCs and GC candidates, including G1, in the outer
halo of M31. We compared the photometry of G1 in the $B$ and $V$ bands
between these two studies; the results match closely. In this paper,
we adopt the CCD $BVR$ photometry of \cite{rhh94}, and the
photographic $U$-band photometry of \cite{Sidney69}, with a
photometric uncertainty of 0.08 mag as suggested by
\cite{gall04}. Based on {\sl HST} images, \cite{bh01} detected and
published photometry for 114 GC candidates associated with M31,
including 32 new objects. Their $V$-band photometry is in good
agreement with that of \cite{Sidney69} and \cite{rhh94}, although
they do not provide their photometric uncertainties. However,
\cite{bh01} compared their {\sl HST} photometry with the ground-based
measurements compiled by \cite{bh00}, and found that the median
offset in $I$ is $0.06\pm0.04$ mag. Therefore, we adopt $0.06$ mag
as the photometric uncertainty in the $I$ band. Using the Two Micron
All Sky Survey (2MASS) database, \cite{gall04} identified 693 known
and candidate GCs in M31, and listed their 2MASS $JHK_{\rm s}$
magnitudes. \cite{gall04} transformed all 2MASS magnitudes to the CIT
photometric system (Elias et al. 1982, 1983) using the color
transformations in \cite{Carpenter01}. However, we need the original
2MASS $JHK_{\rm s}$ magnitudes to compare our observational SEDs with
the SSP models, so we reversed this transformation using the same
procedures. Since \cite{gall04} do not provide the photometric
uncertainties in $JHK_{\rm s}$, we obtained these by comparing the
magnitudes with fig. 2 of \cite{Carpenteretal01}, where the observed
photometric rms uncertainties in the time series are shown as a
function of magnitude, for stars brighter than the observational
completeness limits. In fact, the photometric uncertainties adopted do
not affect our results significantly, as we showed in \cite{fan06}
(see their section 4.3 for details). The full set of ultraviolet,
optical broad-band, and 2MASS NIR photometry of G1 is listed in Table
1. The $UBVRI$ and 2MASS magnitudes are given in the Vega system
\cite{schneider02}. For convenience, we converted all observational
magnitudes to the AB system, following the procedures recommended in
BC03.

\begin{table}[]
\begin{center}
\caption[]{Ultraviolet, optical broad-band, and NIR 2MASS photometry of
G1.} \label{tab correlation}
\begin{tabular}{lcc}
\hline\noalign{\smallskip}
Filter & Magnitude (uncertainty) & Reference \\
\hline\noalign{\smallskip}
   FUV   & 18.972 (0.031) & \cite{rey06}\\
   NUV   & 18.014 (0.012) &              \\
         &                &              \\
   $U$   & 14.85  (0.08)  & \cite{Sidney69}\\
         &                &              \\
   $B$   & 14.584 (0.013) & \cite{rhh94}\\
   $V$   & 13.750 (0.007) &\\
   $R$   & 13.191 (0.010) &\\
         &                &              \\
   $I$   & 12.684 (0.060) &  \cite{bh01}\\
         &                &              \\
   $J$   & 11.858 (0.054) &  \cite{gall04}\\
   $H$   & 11.127 (0.054) &\\
$K_{\rm s}$ & 11.016 (0.054) &\\
\noalign{\smallskip}\hline
\end{tabular}
\end{center}
\end{table}

\subsection{Reddening and metallicity}

To obtain the intrinsic SED of G1, the photometric data must first be
dereddened. Since G1 is located in the halo of M31, i.e., far away
from the galaxy's disk, it is (for all practical purposes) only
affected by Galactic foreground extinction. In fact, some authors have
demonstrated that G1 is affected by a negligible amount of
reddening. \cite{meylan01} used {\sl HST}/WFPC2 observations in the
F555W and F814W filters, and applied \cite{sarajedini94}'s method to
simultaneously determine the cluster's reddening and metallicity; they
obtained a reddening of $E(V-I)=0.05\pm0.02$ mag toward G1, which is
less than the Galactic foreground extinction. \cite{Sidney69}
studied the reddening in the halo of M31 by comparing the colors of
clusters with the same line-strength index in the Galaxy and in M31,
and obtained a mean reddening of $E(B-V)=0.08\pm0.02$ mag for the
clusters in the halo of M31. \cite{bh00} determined the reddening
for each individual cluster using correlations between optical and
infrared colors and metallicity, and by defining various
`reddening-free' parameters based on their large database of
multi-color photometry. For G1, Barmby et al. (2000, also P. Barmby,
priv. comm.) obtained $E(B-V)=0.09\pm0.02$ mag. In this paper,
we adopt the reddening value from \cite{bh00}. The values for the
extinction coefficient, $R_{\lambda}$, were obtained by interpolating
the interstellar extinction curve of \cite{car89}.

Cluster SEDs are determined by the combination of their ages and
metallicities, which is often referred to as the age-metallicity
degeneracy. Therefore, the age of a cluster can only be constrained
accurately if the metallicity is known with confidence, from
independent determinations. There exist two metallicity determinations
for G1. \cite{hbk91} derived metallicities for 150 M31 GCs, including
G1, using the strengths of six absorption features in the clusters'
integrated spectra. The resulting metallicity of G1 is $\rm
[Fe/H]=-1.08\pm0.09$.  \cite{meylan01} used {\sl HST}/WFPC2
photometry to construct deep CMDs for G1, combined with the shape of
the red-giant branch as calibrated by \cite{sarajedini00}, to derive
the mean metallicity of G1 on the scale of \cite{zinn84}, $\rm
[Fe/H]=-0.95\pm0.09$. In this paper, we adopt $\rm
[Fe/H]=-1.08\pm0.09$ for G1.

\section{The stellar population of G1}

\subsection{Stellar populations and synthetic photometry}

In this section, we compare the SED of G1 with theoretical stellar
population synthesis models. We start by using the BC03 SSP models,
which have been upgraded from the earlier \cite{bc93,bc96} versions,
and now provide the evolution of the spectra and photometric
properties for a wider range of stellar metallicities.  BC03 provide
26 SSP models (both of high and low spectral resolution) using the
Padova-1994 evolutionary tracks, half of which were computed based
on the \cite{salp55} IMF with lower and upper-mass cut-offs of
$m_{\rm L}=0.1~M_{\odot}$ and $m_{\rm U}=100~M_{\odot}$,
respectively. The other thirteen were computed using the
\cite{chabrier03} IMF with the same mass cut-offs. In addition, BC03
provide 26 SSP models using the Padova-2000 evolutionary tracks. In
this paper, we will use all of these SSP models to determine the
most appropriate age and mass for G1. These SSP models contain 221
spectra describing the spectral evolution of SSPs from
$1.0\times10^5$ yr to 20 Gyr. The evolving spectra include the
contribution of the stellar component at wavelengths from 91\AA~~to
$160\mu$m.

Since our observational data are integrated luminosities through a
given set of filters, we convolved the theoretical SSP SEDs of BC03
with the FUV and NUV, broad-band $UBVRI$, and 2MASS $JHK_{\rm s}$
filter response curves to obtain synthetic ultraviolet, optical, and
NIR photometry for comparison. The synthetic magnitude in the AB
magnitude system for the $i{\rm th}$ filter can be computed as

\begin{equation}
m_{\lambda_i}=-2.5\log\frac{\int_{\lambda}F_{\lambda}\varphi_{i}
(\lambda){\rm d}\lambda}{\int_{\lambda}\varphi_{i}(\lambda){\rm
d}\lambda}-48.60,
\end{equation}
where $F_{\lambda}$ is the theoretical SED and $\varphi_{i}$ the
response curve of the $i{\rm th}$ filter. $F_{\lambda}$ varies as a
function of age and metallicity.

\subsection{Fit results}

We use a $\chi^2$ minimization approach to examine which SSP models
are most compatible with the observed SEDs, following

\begin{equation}
\chi^2=\sum_{i=1}^{10}{\frac{[m_{\lambda_i}^{\rm
intr}-m_{\lambda_i}^{\rm mod}(t)]^2}{\sigma_{i}^{2}}},
\end{equation}
where $m_{\lambda_i}^{\rm mod}(t)$ is the integrated magnitude in
the $i{\rm th}$ filter of a theoretical SSP at age $t$,
$m_{\lambda_i}^{\rm intr}$ represents the intrinsic integrated
magnitude in the same filter, and

\begin{equation}
\sigma_i^{2}=\sigma_{{\rm obs},i}^{2}+\sigma_{{\rm mod},i}^{2}.
\end{equation}
Here, $\sigma_{{\rm obs},i}^{2}$ is the observational uncertainty, and
$\sigma_{{\rm mod},i}^{2}$ is the uncertainty associated with the
model itself, for the $i{\rm th}$ filter. \cite{charlot96} estimated
the uncertainty associated with the term $\sigma_{{\rm mod},i}^{2}$ by
comparing the colors obtained from different stellar evolutionary
tracks and spectral libraries. Following \cite{wu05}, we adopt
$\sigma_{{\rm mod},i}^{2}=0.05$.

The BC03 SSP models based on the Padova-1994 evolutionary tracks
include six initial metallicities, $Z= 0.0001, 0.0004, 0.004, 0.008,
0.02\, (Z_\odot)$, and 0.05, corresponding to ${\rm [Fe/H]}=-2.25,
-1.65, -0.64, -0.33, +0.09$, and $+0.56$. However, the BC03 SSP
models based on the Padova-2000 evolutionary tracks include six
partially different initial metallicities, $Z = 0.0004$, 0.001,
0.004, 0.008, 0.019 $(Z_\odot)$, and 0.03, i.e., ${\rm
[Fe/H]}=-1.65, -1.25, -0.64, -0.33, +0.07$, and $+0.29$.  Spectra
for other metallicities can, in principle, be obtained by
interpolation of the appropriate spectra for any of these
metallicities, although this is not necessarily advisable or
straightforward (Frayn \& Gilmore 2002). Instead, we adopt the most
appropriate model metallicity for the analysis performed in this
paper. Since we have good estimates of the metallicity and reddening
values of G1 (see \S 2.3), the cluster age is therefore the sole
parameter to be estimated (for a given IMF and extinction law, both
of which we assume to be universal).

None of the SSP models fit the photometric data point in the {\sl
GALEX} FUV band as well as the other nine data points. (We checked
that the image of G1 in the FUV band is not affected by instrumental
problems.) Given that G1 contains an old stellar population, the
most likely physical explanation for this FUV excess compared to the
`standard' BC03 SSP models is the presence of a significant number
of FUV-bright, hot, extreme horizontal-branch (EHB) stars giving
rise to the well-known `ultraviolet upturn' below $\lambda \simeq
2000${\AA} (see, e.g., the review of O'Connell 1999, and references
therein; see also Landsman et al. 1998; Sohn et al. 2006).
(Alternative species, such as AGB-manqu\'e stars or blue stragglers
are expected to be fewer in number in any `normal' stellar
population.)

Since `standard' SSP models do not contain EHB populations, we are
forced to deselect the photometric FUV data point when applying our
fitting routines. In Fig. 1, we show the intrinsic SED of G1 and the
integrated SEDs of the best-fitting models. The dereddened data are
shown as the symbols with error bars (vertical errors for
photometric uncertainties and horizontal error bars for the
approximate wavelength coverage of each filter); open circles
represent the calculated magnitudes of the model SED for each
filter, obtained by convolving the theoretical SSP SEDs with the
appropriate filter response curves. The best reduced-$\chi^2$ values
and ages are listed in Table 2. The mass of G1, also listed in Table
2, can be estimated by comparing the measured luminosity in the $V$
band with the theoretical mass-to-light ($M/L$) ratios. These $M/L$
ratios are a function of the cluster age and metallicity. The
mass-to-light ratios of G1, calculated based on the metallicity
adopted and the age obtained in this paper, are listed in Table 2
for the BC03 SSP models. Based on its present luminosity, $V =
13.750\pm 0.007$ mag, and extinction, $E(B-V) = 0.09$ mag, the
cluster's visual magnitude corrected for the extinction is $V_0 =
13.471\pm 0.007$ mag, assuming a \cite{car89} Galactic reddening law
with $A_V = 0.279$ mag.

(We note that the NUV data point is also marginally affected by the
onset of the UV upturn, which causes a slight mismatch between the
observations and the best-fitting theoretical SSP models.)

\begin{figure*}
\resizebox{\hsize}{!}{\rotatebox{0}{\includegraphics{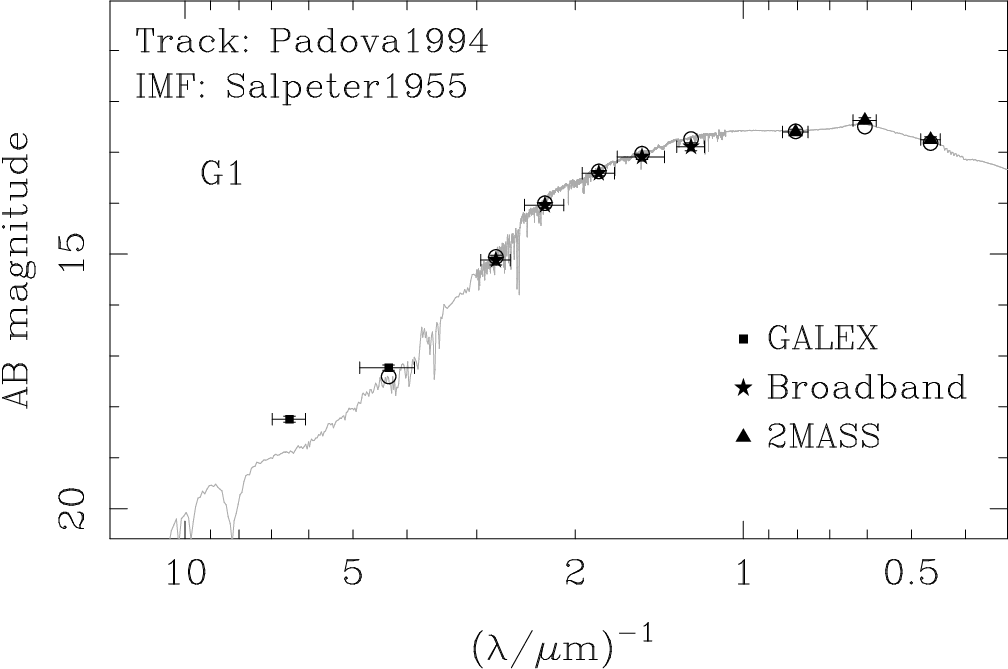}\includegraphics{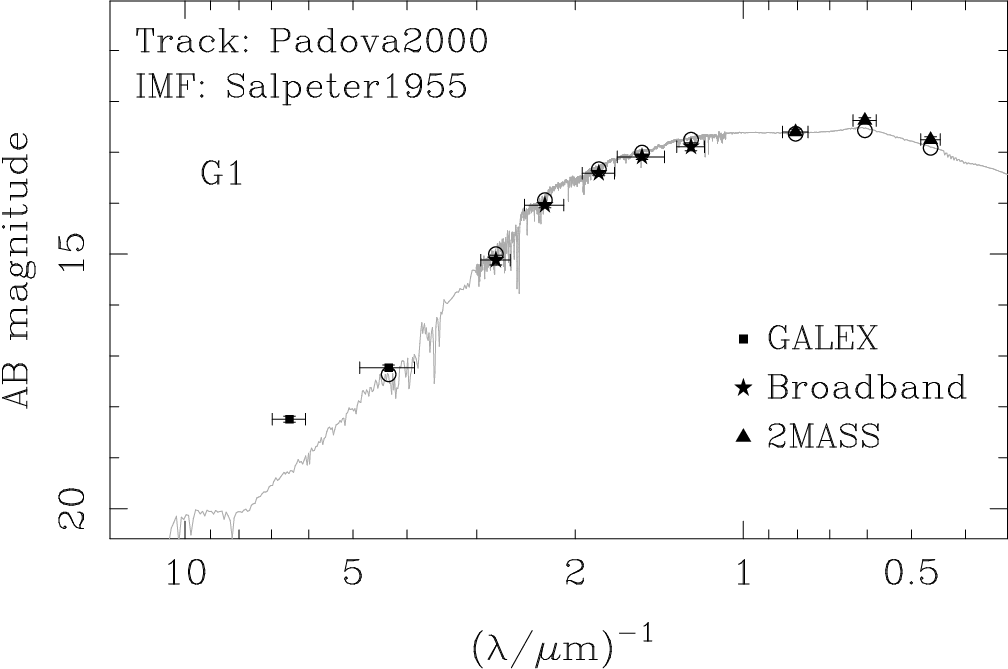}}}
\resizebox{\hsize}{!}{\rotatebox{0}{\includegraphics{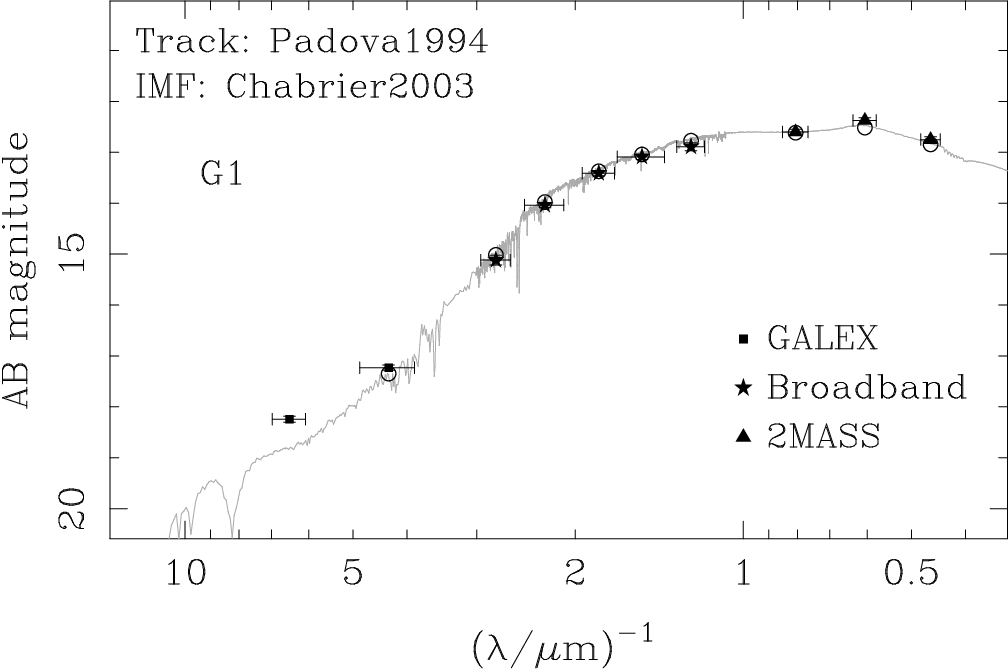}\includegraphics{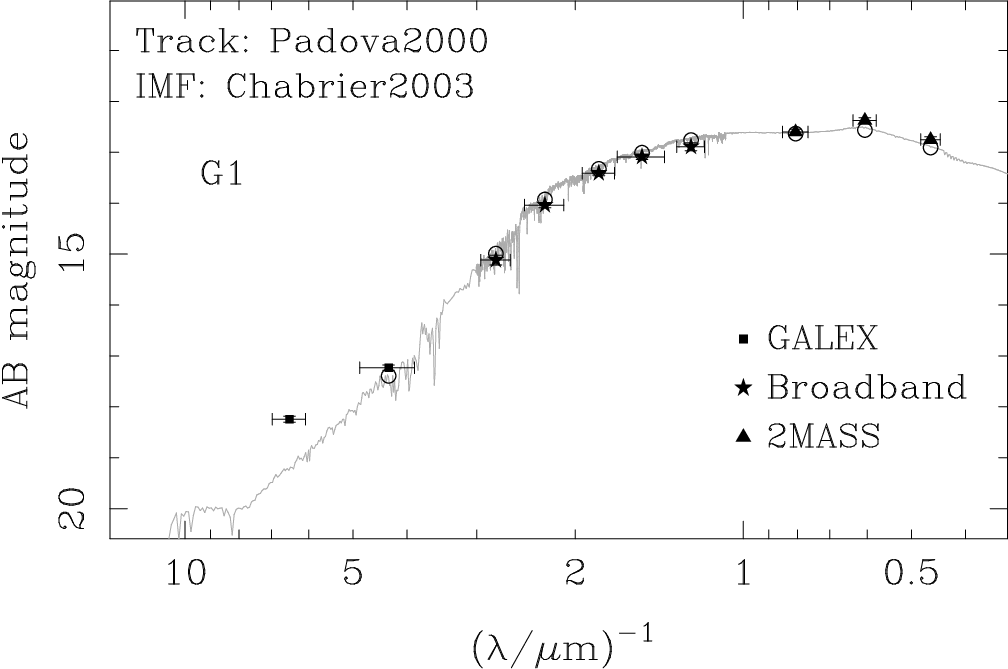}}}
\caption{Best-fitting integrated theoretical BC03 SEDs compared to
the intrinsic SED of G1. The photometric measurements are shown as
the symbols with error bars (vertical for uncertainties and
horizontal for the approximate wavelength coverage of each filter).
Open circles represent the calculated magnitudes of the model SED
for each filter. We did not use the FUV photometric data point for
the fits (see text).} \label{fig1}
\end{figure*}

\subsection{Age and mass}

In the previous section we determined the best-fitting age and mass
of G1 based on different theoretical SSP models. From Table 2 we
conclude that, within the errors, the ages obtained from the
different BC03 models are internally consistent. The mean age of G1
is $18.23\pm{1.76}$~Gyr. This is in excellent agreement with the
only other available (rough) age estimate for the cluster by Meylan
et al. (2001), who estimated its age to be $\sim$15 Gyr. However, we
note that the age of G1 obtained in this paper is older than the
current-best estimate of the age of the Universe, of order 13.7 Gyr,
as discussed in \S 3.1. We will discuss this problem in \S 4.

\begin{table}[]
\begin{center}
\caption[]{Age and mass estimates of G1 based on the BC03 models.}
\label{tab correlation}
\begin{tabular}{cccccc}
\hline\noalign{\smallskip}
Evolutionary Track &     IMF      &      Age         & $\chi^2/\rm{degree~~of~~freedom}$        & $M/L_V$    &    Mass      \\
&                  &     (Gyr)    &                  & $(M/L_V)_\odot$ &  $(10^7 M_\odot)$ \\
\hline\noalign{\smallskip}
Padova 1994 &  \cite{salp55}     & $19.68\pm{0.75}$ & 3.04     &  5.10     &   $1.06\pm{0.07}$ \\
                   &                     &                 &           &           &         \\
Padova 1994 &  \cite{chabrier03} & $19.79\pm{0.50}$ & 2.69     &  3.15     &    $0.65\pm{0.04}$ \\
                   &                     &                 &           &           &             \\
Padova 2000 &  \cite{salp55}     & $15.44\pm{0.78}$ & 5.36     &  4.14     &    $0.86\pm{0.05}$ \\
                   &                     &                 &           &           &           \\
Padova 2000 &  \cite{chabrier03} & $18.01\pm{2.00}$ & 5.27     &  2.79     &     $0.58\pm{0.04}$ \\
\noalign{\smallskip}\hline
\end{tabular}
\end{center}
\end{table}

We conclude that the various mass estimates listed in Table 2 place
G1 firmly at the top of the cluster mass function in the Local
Group. \cite{meylan01} presented three estimates of the total mass
of G1, (i) a King-model mass (King 1966) of $1.5\times 10^7
M_\odot$, (ii) a virial mass of $0.75\times 10^7~M_\odot$; and (iii)
a mass based on a King-Michie model (as defined by Gunn \& Griffi 1979)
fitted simultaneously to the surface brightness profile and the
central velocity dispersion value, estimated between $1.4\times 10^7
M_\odot$ and $1.7\times 10^7 M_\odot$. Our results are in reasonable
agreement with \cite{meylan01}, although we are aware that the King
and King-Michie mass estimates of \cite{meylan01} are up to a
factor of two greater than our photometric mass estimates. This is
not too surprising in view of the model assumptions made.
\cite{cohen06} recently obtained an optical velocity dispersion for
the cluster using the Keck/HIRES spectrograph, and derived an
aperture-corrected line-of-sight velocity dispersion, $\sigma_{\rm
los} = 25.5 \pm 1.5$ km s$^{-1}$ (where we have averaged the values
she obtained for the two reddest wavelength ranges analyzed; see
also Djorgovski et al. 2002). We recently redetermined a projected
half-light radius for G1 of $r_{\rm h} = 6.5 \pm 0.3$ pc (Ma et al.
2007b). Thus, based on these most recent results, the dynamical
(virial) mass of G1 is $M_{\rm vir} = (7.37 \pm 2.15) \times 10^6
M_\odot$. This is in excellent agreement with the photometric mass
estimates obtained in this paper. In turn, this strongly supports
the notion that G1 must have had a close-to-`normal' stellar IMF, in
order for it to have survived dissolution due to internal two-body
relaxation until the present time (see also Ma et al. 2006). In
particular, this is driven by the observation that if the IMF is too
shallow, i.e., if a cluster is significantly depleted in low-mass
stars compared to (for instance) the solar neighborhood, it will
disperse within a few orbital periods around its host galaxy's
center, and most likely within about a Gyr of its formation (e.g.,
Gnedin \& Ostriker 1997; Goodwin 1997; Smith \& Gallagher 2001;
Mengel et al. 2002; Rose, Kouwenhoven \& de Grijs, in prep.).

From the recent work of \cite{ma06}, the intrinsically most luminous
M31 GC, 037-B327, has been suggested to be the most massive GC in the
Local Group, with a total mass of $\sim (3.0 \pm 0.5) \times 10^7
M_\odot$, also determined photometrically and somewhat depending on
the SSP models used, the metallicity and age adopted, and the IMF
representation. However, \cite{cohen06} pointed out that the
photometric mass of this cluster had likely been overestimated due to
an incorrect extinction correction.  Nevertheless, she also confirmed
the nature of 037-B327 as one of the most massive GCs in the Local
Group, with a dynamical mass similar to that of G1.  It is intriguing
that these two most massive GC in M31 both are significantly more
massive than the most massive Galactic GC, $\omega$ Cen [$\sim (2.9 -
5.1) \times 10^6 M_\odot$; \cite{meylan02}].

In fact, the high mass of these clusters raises additional,
intriguing questions regarding the nature of these objects in
general, and of G1 in particular (see also Federici et al. 2007; Ma
et al. 2007b, and references therein). It has been speculated that
these objects may be nucleated dwarf galaxies instead of genuine
GCs. In support of this notion, we point out that Gieles et al.
(2006) suggest that there may be a physical upper limit to the mass
of a star cluster that is not merely the result of size-of-sample
effects. This maximum mass depends to some extent on the galactic
environment; for their example galaxies, M51 and the Antennae
system, they find a physical upper limit to the stellar mass of
$\sim (10^5 - 10^6) M_\odot$. Our values derived for both the
photometric and the virial mass of G1 are well above these suggested
upper mass limits. This may, therefore, provide an additional
(although circumstantial) proverbial nail in the coffin of G1 as a
normal GC.

\subsection{Luminosity of the hot, extreme horizontal-branch stars}

As discussed in \S 3.3, none of the BC03 SSP models fit the
photometric data point in the {\sl GALEX} FUV band as well as the
other nine data points. EHB stars may be responsible for this excess
in FUV band. In fact, \cite{rich96} found some bluer
horizontal-branch stars extending to $(V-I)=0.0$ mag, based on their
{\sl HST}/WFPC2 observations. In this section, we will calculate the
luminosity of the EHB stars. We assume that the excess in the {\sl
GALEX} FUV band is solely due to these EHB stars. The magnitude
differences between the four SSP models employed in this paper and the
photometric data points are 0.63, 1.00, 0.56, and 0.94 mag,
respectively, corresponding $(2.44, 1.07, 2.11, \mbox{ and }
1.00)\times10^2 L_\odot$, respectively, with a mean number of EHB
stars in G1 of $(1.65\pm0.63)\times10^2 L_\odot$.

\subsection{Comparison between G1 and S312}

\cite{brown04} showed that a 10 Gyr old population in M31 has a
main-sequence turnoff at about $m_{\rm F814W} = 28.8$ mag. Their deep
observations needed exposures of 39.1 hours in F606W and 45.5 hours
F814W, spanning 120 orbits of {\sl HST}/ACS imaging observations
(Brown et al. 2003, 2004). In the future, such deep {\sl HST}
observations will only be obtained for a very small number of fields
(e.g., Rich et al. 2005). As discussed in \S 1, \cite{ma07a}
constrained the age of the M31 GC S312 by comparing multicolor
photometry and theoretical stellar population synthesis models. It is
encouraging that the age obtained by \cite{ma07a} is in good
agreement with the previous determination based on main-sequence
photometry (Brown et al. 2004), i.e., $9.5^{+1.15}_{-0.99}$ Gyr versus
$10^{+2.5}_{-1}$ Gyr. S312 is one of the few extragalactic GCs for
which the age can be determined from main-sequence photometry. By
comparing the ages of S312 and G1 determined using the same method, we
can conclude that S312 is younger than the majority of the Galactic
GCs at the same metallicity, and G1 is as old as the oldest Galactic
GCs. In fact, if we try to fit the intrinsic SEDs of G1 by the
theoretical BC03 SSP SEDs at an age of 10 Gyr, the resulting fit is
very poor indeed, particularly in the ultraviolet. Therefore, we
conclude that the method used in this paper, by which the ages of both
S312 and G1 have been determined, can be used to determine the ages of
old stellar populations to satisfactory precision, and in particular
to distinguish between young and old populations.

\section{Summary and Conclusions}

In this paper, we first determined the age and mass of the M31 GC G1,
as well as the realistic uncertainties associated with these
estimates, by comparing its multicolor photometry with theoretical
stellar population synthesis models. Our multicolor photometric data
were obtained from {\sl GALEX} FUV and NUV, broad-band optical
$UBVRI$, and 2MASS $JHK_{\rm s}$ observations, which form an SED
covering the wavelength range from 2267 to 20,000{\AA}. Our results
confirm that G1 is one of the oldest and most massive GCs in the Local
Group -- that is, if it is indeed a genuine GC given that its mass is
well in excess of the physical maximum mass predicted by the models of
Gieles et al. (2006).

The age and mass obtained in this paper are somewhat dependent on
the SSP model adopted. It is evident that the age of
$18.23\pm{1.76}$~Gyr for G1 based on the BC03 models is greater than
the currently accepted age of the Universe. However, we must keep in
mind that the BC03 SSP models were calculated for ages up to 20 Gyr.
In fact, ages derived for objects such as GCs and galaxies in excess
of that of the Universe only mean that these objects are among the
oldest objects in the Universe.

In the context of the BC03 models and their associated age range up to
20 Gyr, our derived age for G1 is consistent with the suggestion by
the {\sl WMAP} team that the oldest GCs may have formed within the
first 1.7 Gyr after the Big Bang (see \S 1).

The integrated FUV flux depends mainly on the fractional number of
horizontal-branch (HB) stars with temperatures hotter than $T_e \sim
10,000$ K, with a modest dependence on their temperature distribution
(see Rey et al. 2007 and references therein). Older GCs produce more of
these hot HB stars and they are thus more likely to produce stronger
FUV fluxes at a given metallicity (see Rey et al. 2007 and references therein)
\cite{lee03} showed that the addition of FUV
photometry to optical data can discriminate cleanly among young ($<$1
Gyr), intermediate-age (3--5 Gyr), and old ($>$ 14 Gyr) GCs. Young and
very old GCs exhibit a significant FUV-to-optical spectral continuum
slope, but intermediate-age clusters are relatively faint in the FUV
(see fig. 2 of Lee et al. 2003). Figure 6 of \cite{rey06} implies that
the age of G1 may be similar to that of the oldest GCs.

Overall, we therefore conclude that G1 is indeed among the oldest and
most massive building blocks of M31, and provides a key limitation to
the age of the Universe, although we caution that our results also
provide circumstantial support to the suggestion that the cluster may
not be a genuine GC.

\section*{Acknowledgments}
We are indebted to the referee for thoughtful comments and
insightful suggestions that improved this paper greatly. This work
has been supported by the Chinese National Natural Science
Foundation through Grant Nos. 10873016, 10803007, 10473012,
10573020, 10633020, 10673012, and 10603006; and by National Basic
Research Program of China (973 Program) No. 2007CB815403. RdG
acknowledges partial financial support from the Royal Society in the
form of a UK-China International Joint Project. SCR acknowledges
partial support from KOSEF through the Astrophysical Research Center
for the Structure and Evolution of the Cosmos (ARCSEC). This paper
makes use of data from the Two Micron All Sky Survey, which is a
joint project of the University of Massachusetts and the Infrared
Processing and Analysis Center, funded by NASA and the National
Science Foundation. This paper is also partially based on archival
observations with the NASA/ESA {\sl Hubble Space Telescope},
obtained at the Space Telescope Science Institute (STScI), which is
operated by the Association of Universities for Research in
Astronomy, Inc., under NASA contract NAS 5-26555. This research has
made use of NASA's Astrophysics Data System Abstract Service. This
reasearch is partially based on archival data from the NASA GALEX
mission developed in cooperation with the Centre National d'Etudes
Spatiales of France and the Korean Ministry of Science and
Technology.

\end{document}